\documentclass[utf8]{FrontiersinHarvard} % for articles in journals using the Harvard Referencing Style (Author-Date), for Frontiers Reference Styles by Journal: https://zendesk.frontiersin.org/hc/en-us/articles/360017860337-Frontiers-Reference-Styles-by-Journal
\usepackage{url,hyperref,lineno,microtype,subcaption}
\usepackage[onehalfspacing]{setspace}

\usepackage[T1]{fontenc}
\usepackage{lmodern}

\def\keyFont{\fontsize{8}{11}\helveticabold }
\def\firstAuthorLast{Jahani {et~al.}} %use et al only if is more than 1 author
\def\Authors{Bahar Jahani\,$^{1}$, Matsanga Leyila Kaseka\,$^{2,3}$, Marta Kersten-Oertel\,$^{1}$, \\and Yiming Xiao\,$^{1,*}$}
\def\Address{$^{1}$Department of Computer Science and Software Engineering, Concordia University, Montreal, Quebec, Canada \\
$^{2}$Pediatric Neurology, CHU Sainte-Justine, Montreal, Quebec, Canada \\
$^{3}$Department of Neurosciences, University of Montreal, Montreal, Quebec, Canada.  }
\def\corrAuthor{Yiming Xiao}

\def\corrEmail{yiming.xiao@concordia.ca}

\begin{document}
\onecolumn
\firstpage{1}

\title[NeuroVase: Mobile AR for Clinical Stroke Education]{NeuroVase: A Tangible Mobile Augmented Reality Learning System for Neurovascular Anatomy and Stroke Education} 

\author[\firstAuthorLast ]{\Authors} %This field will be automatically populated
\address{} %This field will be automatically populated
\correspondance{} %This field will be automatically populated

\extraAuth{}% If there are more than 1 corresponding author, comment this line and uncomment the next one.
%\extraAuth{corresponding Author2 \\ Laboratory X2, Institute X2, Department X2, Organization X2, Street X2, City X2 , State XX2 (only USA, Canada and Australia), Zip Code2, X2 Country X2, email2@uni2.edu}

\maketitle

\begin{abstract}
Stroke remains a leading cause of mortality and disability worldwide, requiring rapid and informed clinical decision-making. A solid spatial understanding of cerebrovascular anatomy and vascular territories in relation to stroke symptoms and severity is critical for timely clinical decision and patient care. However, this knowledge is typically conveyed through static 2D diagrams and printed materials, which can hinder mastery of the complex neurovascular system and their clinical implications. Mobile augmented reality (AR) offers an accessible medium for delivering intuitive 3D anatomical education, yet applications focused on the neurovascular system and stroke remain limited despite the demand. To address this, we propose \textit{NeuroVase}, a tablet-based mobile AR platform within a structured pedagogical framework that enhances stroke-related neuroanatomy learning by providing an interactive, engaging, and accessible alternative to traditional methods. \textit{NeuroVase} features a dual-mode setup, using tangible cue cards as standalone study aids while also serving as interactive markers for AR content delivery. A custom learning curriculum focused on cerebrovascular anatomy and stroke supports exploration of vascular territories, stroke syndromes, and arterial occlusions, in the context of annotated 3D anatomical models in \textit{NeuroVase}. A controlled user study with 40 participants revealed that \textit{NeuroVase} is an effective and user-friendly AR platform to facilitate complex anatomical and physiological education, compared with traditional learning.

%%% Leave the Abstract empty if your article does not require one, please see the Summary Table for full details.
%\section{}
%For full guidelines regarding your manuscript please refer to \href{http://www.frontiersin.org/about/AuthorGuidelines}{Author Guidelines}.

%As a primary goal, the abstract should render the general significance and conceptual advance of the work clearly accessible to a broad readership. References should not be cited in the abstract. Leave the Abstract empty if your article does not require one, please see \href{http://www.frontiersin.org/about/AuthorGuidelines#SummaryTable}{Summary Table} for details according to article type. 

\tiny
 \keyFont{ \section{Keywords:} Mobile augmented reality, medical education, vasculature, stroke, neuroanatomy} %All article types: you may provide up to 8 keywords; at least 5 are mandatory.
\end{abstract}

\section{Introduction}
The study of neuroanatomy and its associated brain physiology is fundamental to both the investigation and clinical management of neurological disorders. However, mastering this material is challenging due to the brain’s intricate spatial organization and the complexity of its functional relationships \citep{giles2010clinical}. These challenges are particularly evident in time-sensitive clinical contexts such as stroke care, where accurate and timely interpretation of clinical imaging (e.g., CT scans) and correct assessment of stroke lesion location and extent in relation to cerebrovascular anatomy are critical for treatment decision-making. Yet, currently there is an urgent need for effective learning content creation and delivery for stroke education \citep{Kaiz2025}. Traditional educational methods, which rely on 2D illustrations and pamphlets do not effectively convey the brain’s three-dimensional complexity. In contrast, virtual and augmented reality (VR/AR) technologies offer promising alternatives, combining their increasing accessibility with powerful capabilities for interactive, three-dimensional visualization. These technologies have started to show their benefit and being adopted for anatomy education~\citep{stirling2020use, kolecki2022assessment, stojanovska2020mixed, veer2022incorporating, zimmer2024perspective}, including neuroanatomy~\citep{intro_neuro, chen2020can, maniam2020exploration, aridan2024neuroanatomy, gurses2024creating}.

With growing evidence that VR and AR are effective tools for learning, numerous software platforms and studies have explored their use in neuroanatomy and brain function education. With virtual reality, \cite{Sonia} designed a novel software system that enhances neuroanatomy education by offering immersive and customizable visualization of complex 3D brain structures and neural pathways. \cite{lopez2021} found that VR-facilitated interactive learning helped participants perform better in identifying brain structures and describing their functions compared to traditional lectures. Additionally, \cite{souza2021} introduced a VR-based anatomical puzzle game, which received positive subjective evaluations, but failed to show a clear improvement in knowledge acquisition and retention. In contrast, \cite{gloy2022} demonstrated that VR-based approaches could enhance learning efficiency and retention compared to standard atlas textbooks. \cite{gurses2024creating} evaluated VR- and AR-based neuroanatomy educational models developed using photogrammetry for neurosurgery residents and medical students. Results showed significant improvements in knowledge scores post-training and positive user feedback, indicating that VR/AR models can effectively enhance neuroanatomy education as a valuable supplement to traditional methods. Similarly, \cite{aridan2024neuroanatomy} conducted a study using photogrammetry-based 3D models of real human brains, finding that students learning via VR exhibited enhanced spatial understanding and higher satisfaction compared to those using 3D printed models or traditional read-only materials. Finally, \cite{mcnaughton2025design}  introduced a VR application that provides an interactive 3D brain model with integrated MRI data to enhance neuroanatomy education and support students’ spatial understanding and MRI interpretation. Preliminary user feedback showed that the app is intuitive and engaging, with potential to improve learning outcomes in neuroanatomy and MRI reading skills.

In AR applications, \cite{StrokeEDAR} developed a tablet-based AR system to display general brain lobes and textual learning content for stroke education, which proved more effective than traditional pamphlets in explaining stroke concepts. Furthermore, \cite{Moro} also compared tablet-based AR, headset-based AR, and VR in teaching anatomy, and found no significant differences in learning effectiveness among these methods. \cite{GreyMapp} developed a mobile AR application for neuroanatomy education, which demonstrated reduced cognitive load and increased motivation for learning, but did not outperform traditional methods in terms of learning outcomes. Similarly, \cite{Mendez-Lopez2022} found that students prefer mobile AR-based anatomical learning than learning with textbook coloring, one popular anatomical education technique while both achieved similar knowledge gain. Using an AR headset, \cite{WhiteMatter} proposed virtual white matter fiber dissection to teach key neural pathways, showing promising potential for both educational and clinical use. \cite{zhao2024first} evaluated the feasibility and effectiveness of using AR with the Magic Leap headset for neuroanatomy education with second-year medical students, combining 3D virtual brain models, MRI data, and traditional 2D atlases. 
The results of their study showed that all participants found the AR environment easy to use and a meaningful supplement to conventional teaching, with most reporting increased interest, improved memorization, and high satisfaction. These evidence supports the viability of AR as an engaging educational tool in neuroanatomy and potentially related brain disorders.

Many existing studies have demonstrated positive impacts of AR/VR on neuroanatomy education; however, they often rely on relatively simple 3D brain models focused primarily on general brain lobes, while overlooking related physiological knowledge. In particular, the vascular system and territories that possess strong clinical significance have not been explored. Furthermore, while studies \citep{GreyMapp} suggested complementary roles of AR and paper-based learning, few explored paradigms that integrate them. For example, paper-based learning tools, such as cue cards still remain as the staple instruments in medical education and clinical training. Given their enduring utility and effectiveness, integrating cue cards with mobile AR applications represents a compelling approach to neuroanatomy education.

In this study, we present \textit{NeuroVase}, a tablet-based mobile augmented reality (AR) learning platform designed to support interactive education of vascular neuroanatomy and stroke management, which remain largely under-represented in existing AR/VR educational tools. The system leverages accessible tablet hardware to promote user-friendly and scalable technology-enhanced medical learning. Our work contributes to learning technologies in three key areas. \textit{First}, we developed a clinician-informed pedagogical framework that integrates neuroanatomy, cerebrovascular territories, and stroke-relevant concepts (e.g., stroke location and related symptoms) into a structured AR-supported curriculum, publicly released via \href{https://osf.io/7bq2s/overview?view_only=19cfc4c64ea747d089b841f0fe1ddd32}{\textit{the Open Science Framework (OSF) repository}} to facilitate reproducibility and adoption. \textit{Second}, we introduce a dual-mode learning design that combines mobile AR interaction with tangible cue cards, enabling multi-granularity learning pathways and supporting active, situated learning. \textit{Third}, we propose AR interaction and visualization strategies grounded in embodied learning principles, using physical artifacts to enhance spatial understanding. The educational effectiveness and usability of \textit{NeuroVase} were evaluated through a controlled study with 40 participants, incorporating validated usability measures, semi-quantitative questionnaires, and knowledge assessments, in comparison to traditional paper-based learning.

%% Use \subsection commands to start a subsection.
\section{Methods and Materials}
\label{subsec2}

\begin{figure}[h!]
\begin{center}
\includegraphics[width=1\linewidth]{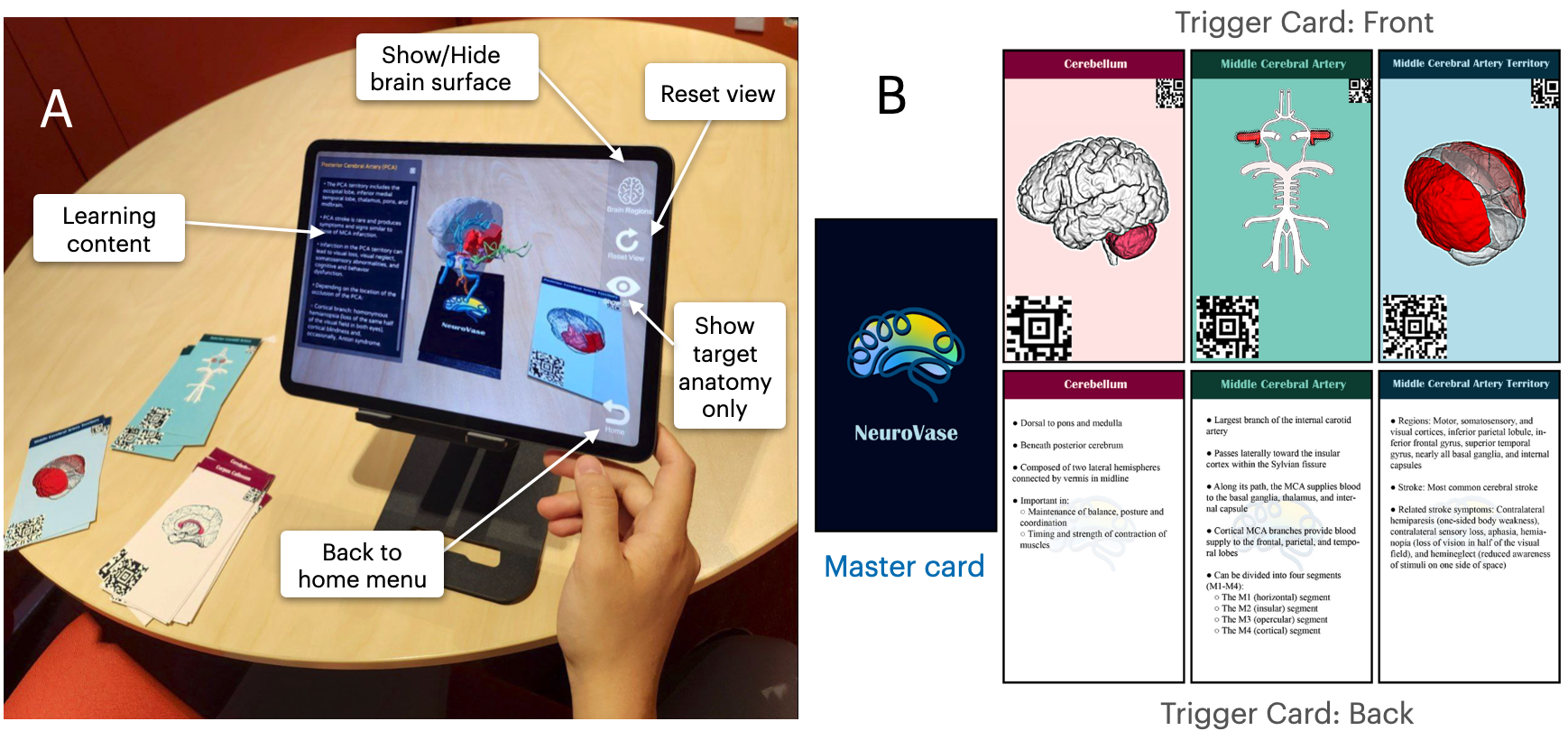}
\end{center}

\caption{Demonstration of AR user interface for \textit{NeuroVase}. A. AR user interface in action showing different elements of the proposed system; B. The master card for virtual anatomy display and three example trigger cards (front and back) to update the learning content in the AR view for the Lobar Anatomy, Vascular Anatomy, and Vascular Territory modules.}
\label{fig:neurovase_overview}
\end{figure}

In the \textit{NeuroVase} system, physical cue cards are used both to interact with the AR content displayed on a tablet and to facilitate offline review of key concepts, as shown in Figure ~\ref{fig:neurovase_overview}. The learning materials include textual explanations paired with the corresponding interactive 3D anatomical models. This study was approved by the Research Ethics Committee of Concordia University. The following sections detail the components of the system.

\subsection{Learning material development}
\label{sec:learningModule}

We developed a learning curriculum in collaboration with a clinical neurologist (co-author M.L.K.) that is specialized in the care of cerebrovascular disorders. The content comprises of three inter-connected modules, including general anatomy of the brain lobes, the cerebral arterial system, and the vascular territories. In the brain lobar module, we introduce the spatial arrangement and basic functions of the general brain divisions, including the frontal, temporal, parietal, and occipital lobes, brainstem (midbrain, pons, and medulla oblongata), and cerebellum. This content provides the foundational knowledge needed for understanding the modules on vascular anatomy and territories, and is especially valuable for users who are new to the topic. For the cerebral arterial system module, we introduce the spatial arrangement, formation, and supplying regions of the main arterial branches, including the internal carotid, anterior cerebral, middle cerebral, posterior cerebral, vertebral, and basilar arteries, as well as the Circle of Willis. Finally, the arterial territory module that was developed based on the stroke population study of \cite{Territories} presents a 3D definition of vascular territories. In this learning module, for each of the anterior cerebral, middle cerebral, posterior cerebral, and vertebral basilar territories, we describe the associated functional brain regions, frequency of stroke occurrence, and related stroke symptoms. To leverage the strengths of each medium, we developed more detailed learning content for the mobile device while keeping a concise summary of the key knowledge points for the cue cards.

\subsection{Data Processing and Digital Model Construction}
\label{sec:data}

For each learning module, we created virtual brain models from different sources. For the learning module on general anatomy of the brain lobes, we extracted the anatomical structures mentioned in the previous section from the BCI-DNI brain atlas by \cite{BCI-DNI}, which includes high-resolution manually labeled brain parcellations from a healthy adult's MRI scan. For the vascular anatomy and territories, we built the digital model from a healthy individual's anatomy. Upon informed consent, the subject was scanned with T1-weighted MRI (1x1x1 $mm^{3}$ resolution) and time-of-flight (TOF) MR angiography (MRA, 0.47x0.47x0.70 $mm^{3}$ resolution) on a 3T MRI scanner. The brain surface and arteries were extracted from the co-registered MRIs with the BEaST algorithm \citep{BEAST} and Frangi vesselness filter \citep{frangi1998multiscale}, respectively. In addition, one co-author, with over 10 years of experience in neuroanatomy also conducted further manual refinement of the arterial labels and segmented the main vessel branches using ITK-SNAP (http://itksnap.org). For the vascular territories, we non-linearly registered the arterial territory atlas by \cite{Territories} to this individual's MRI. This atlas was constructed based on lesion distributions in 1,298 patients with acute stroke. As the atlas has two levels of granularity, we used Level-2 of the subdivisions with four major territories. All segmentations were saved as polygon mesh models in .obj formats for the rendering.  

\subsection{Software Implementation}
\label{sec:dev}
The Unity 3D game engine version 2023.1.16f1 was used for our software application development, and C\# was used for the scripts to communicate with the Unity engine. The Vuforia Engine asset was used to implement the AR-based data display. This asset allows for spatial tracking of a printed image target with a video camera by extracting its image features and then augmenting the virtual content above it. We designed the fronts of the cue cards to be our image targets and provided sufficient geometric features for the Vuforia Engine to detect them. Furthermore, an ``Easy Volume Renderer" asset (https://github.com/mlavik1/UnityVolumeRendering) was used to import the MRI scan into Unity for additional data visualization. 

\subsection{Cue Card Design} 
\label{sec:cards}
The cue cards were made with a common Taro card size (2.75 x 4.75in) and consist of one master card to position the anatomical model spatially in the AR view and a series of trigger cards to update the medical knowledge points and related visual content for the AR display, as well as to serve as an off-line review instrument. Besides the master card, we designed three sets of trigger cards with their respective color-coding: nine cards for the brain lobes, seven for the vascular system, and four for the arterial territories. A few samples of the trigger cards are shown in Figure \ref{fig:neurovase_overview}B. Specifically, each trigger card contains the concise summary of the knowledge points (e.g., an arterial branch) on the back and the related anatomical illustration on the front, which the Vuforia Engine relies on to update the AR content. As Vuforia requires sufficient details (number of features and edges) and enough local contrast for detection, in addition to the anatomical illustration, we also placed two QR codes on the top right and bottom left corners of each card to enhance the robustness for detection. The full set of cue card designs is available at the Open Science Framework (OSF) repository: \href{https://osf.io/7bq2s/overview?view_only=19cfc4c64ea747d089b841f0fe1ddd32}{\textit{https://osf.io/7bq2s}}.

\subsection{User Interaction Paradigms} 
\label{sec:interaction}
Physical cue cards offer an intuitive and tangible way to interact with virtual models, potentially enhancing user engagement and learning. While the master card is used to set the position for the virtual anatomical model, when a user places a trigger card beside the master card in the camera view, the AR display will be updated for the content that the trigger card represents. The user can scale and shift the anatomical model with two-finger pinch/unpinch and swipe, respectively, and rotate the model with one-finger swipe. The ability to manipulate and inspect the digital model from different angles can help enhance the spatial understanding of the target anatomical structures. In addition to finger-gesture-based manipulation, the user can also achieve translation and rotation of the anatomical model by changing the positioning of the master card. To help guide the learning curriculum in the desirable order (lobar anatomy, arterial system, and vascular territories), the cue cards can only be used for the designated learning module, and those for other modules cannot trigger any interface updates. 

\subsection{User Interface Design} 
\label{sec:canvas}
\begin{figure}[h!]
  \begin{center}
\includegraphics[width=1\linewidth]{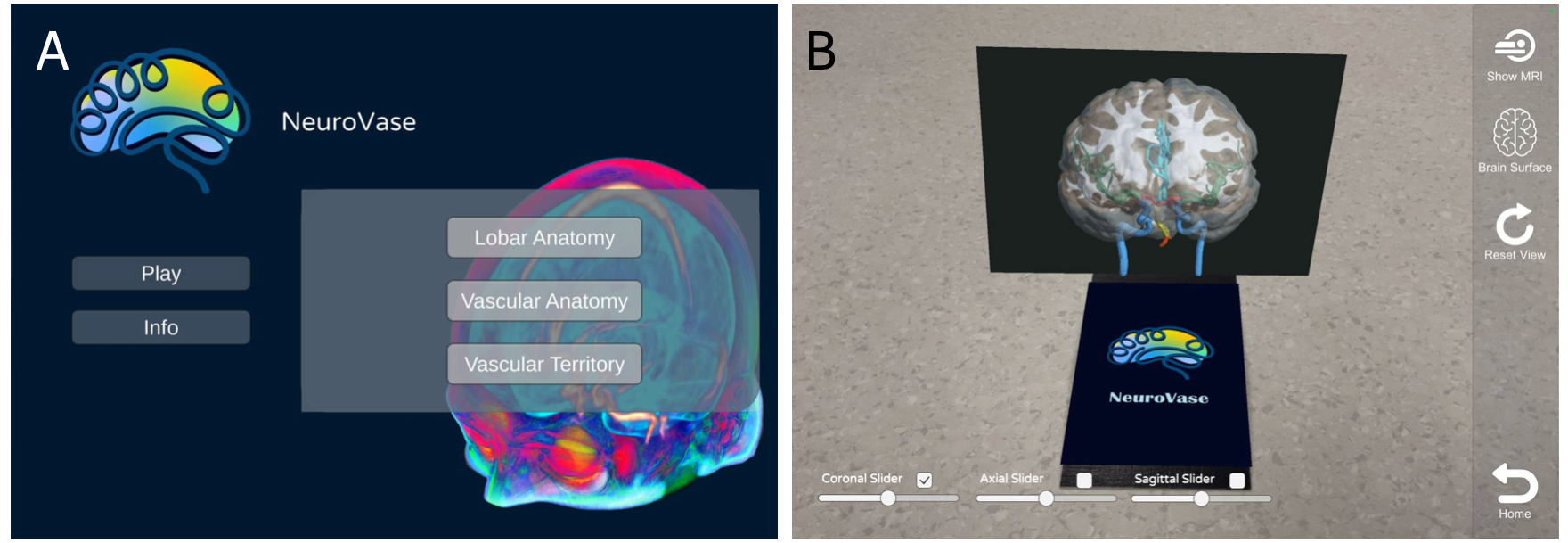}
\end{center}
    \caption{General user interface of \textit{NeuroVase}: A. Demonstration of the landing menu for \textit{NeuroVase}; B. MRI visualization interface for the Vascular Anatomy Module.}
    \label{fig:interfacedemo}
\end{figure}

\begin{figure}[h!]
\begin{center}
\includegraphics[width=1\linewidth]{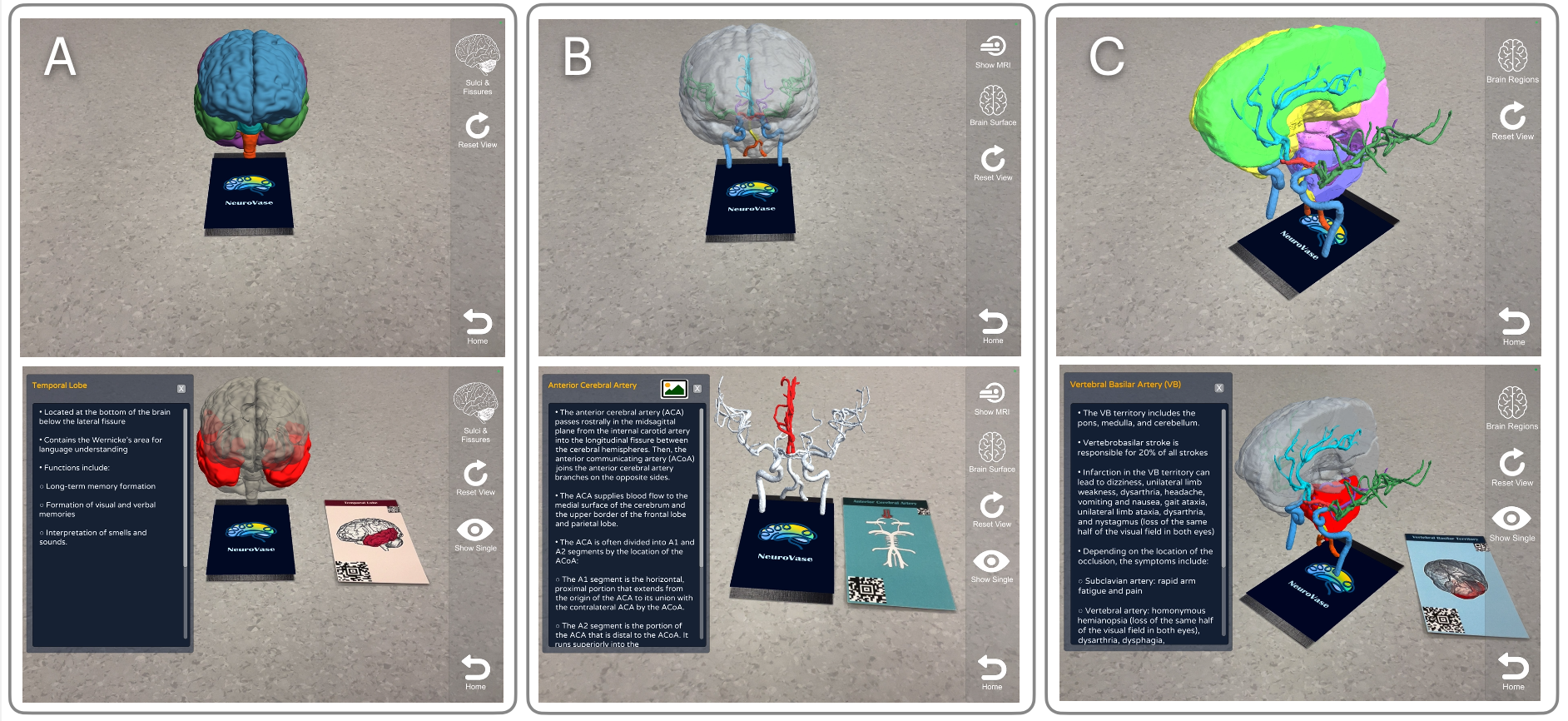}
\end{center}
\caption{Example learning content from the \textit{NeuroVase} interface before and after activation by a cue card, for the Lobar Anatomy module (A), Arterial Anatomy module (B), and Vascular Territory module (C).}
\label{fig:moduleviews}
\end{figure}

We designed an intuitive interface to effectively deliver the learning content. The master menu displays tabs for all three learning modules, as shown in Figure \ref{fig:interfacedemo}A. This modular layout supports flexible content selection and allows for future expansion. Within each module, the interface elements are tailored to its specific learning goals. Figure \ref{fig:moduleviews} illustrates the three learning modules and how the application responds when the master card is detected, and how additional trigger cards reveal relevant anatomical structures and descriptions.

\subsection{User Study Design and System Evaluation}
\label{sec:userStudy}

To assess the usability and effectiveness of the \textit{NeuroVase} system, we conducted a two-prong user study: one with an AR group and another with a control group based on traditional paper-based learning. Upon informed consent, we recruited 40 participants, with 20 for each group.

\subsubsection{AR-Based Learning Group}
For the AR-based learning group, we recruited 20 participants (9 females, 11 males; age = 29.0 ± 4.3 years), all STEM undergraduate or graduate students with limited knowledge of cerebrovascular anatomy and territories. Each participant used the system on a 10.9-inch iPad and completed the designed learning modules within 30 minutes. The user study protocol was as follows. 

First, each participant received a verbal introduction explaining the overall study procedure, followed by a brief hands-on tutorial on the operation of the \textit{NeuroVase} system. They then completed a pre-study quiz consisting of 18 multiple-choice questions to assess their baseline knowledge related to the designed learning material. Next, participants engaged with the AR-based learning modules in a fixed sequence, covering lobar anatomy, arterial anatomy, and vascular territories following the cue cards in a predetermined order. The participants were instructed to interact only with the front side of the cue cards and refrain from reading the content on the back, in order to properly evaluate the iPad-based learning materials. After completing the modules, participants took a post-study quiz identical to the pre-study quiz to measure learning outcomes. Finally, participants were invited to explore the educational content on the backs of the cue cards. At the end of the study, participants completed a questionnaire to evaluate their experience with the \textit{NeuroVase} system. The questionnaire consisted of three parts: the first part was the System Usability Scale (SUS) \citep{brooke1996sus}, a widely recognized tool for assessing software usability. The second part included five customized user experience (UX) questions rated on a 1$\sim$5 Likert scale (1 = strongly disagree, 5 = strongly agree), designed to measure user engagement, enjoyment, perceived usefulness for learning, cue card design quality, and the effectiveness of AR visualization. These questions provided additional insights into the overall user experience. The final part invited participants to share open-ended feedback, including positive aspects, challenges they encountered, and suggestions for improving the system. Lastly, to help better understand the participants' profiles, we also asked the participants to rate their familiarity with AR technology and neuroanatomy, on a scale of 1-5, with 1 being familiar and 5 being unfamiliar.

\subsubsection{Traditional Learning Control Group}
A control group of 20 participants (11 females, 9 males, age = 29.7 ± 5.3 yo), with the same profiles as the AR user group, used traditional paper-based educational materials, instead of the \textit{NeuroVase} AR system. Note that there are no overlapping participants with the AR group. For visual references, instead of 3D models, 2D images from anatomy textbooks were used while the textual content is identical to that in the \textit{NeuroVase} system to ensure fair comparison. The paper-based material was divided into three sections: lobar anatomy, arterial anatomy, and vascular territories, mirroring the AR application's learning curriculum.

Similar to the AR prong of the study, participants received an introduction and completed a pre-study quiz (18 multiple-choice questions). They then studied the provided paper-based materials. Afterward, participants completed a post-study quiz identical to the pre-study to assess learning outcomes. At the study's conclusion, participants answered a user experience questionnaire focused on engagement, enjoyment, and the effectiveness of the learning materials. Unlike the AR group, the System Usability Scale was not included. Finally, participants provided comments on the strengths and weaknesses of the paper-based approach. We asked participants to rate their familiarity with neuroanatomy on a scale of 1$\sim$5 (1 = familiar, 5 = unfamiliar).

\subsection{Data analysis} 
For the pre- and post-study quizzes, we scored them out of 100\%, with each question weighted equally, and compared the scores using the Mann-Whitney U test to confirm the learning gain among the participants within and between groups. The SUS consists of 10 questions rated on a scale of 1-5 (strongly disagree to strongly agree), evaluating system complexity, ease of use, and confidence when using a software system. For each of the scores of the 10 questions $x$, odd-numbered questions are scored as $x-1$ while even-numbered questions are scored as $5-x$. The total score is then summed and multiplied by 2.5 to yield the final score, with a maximum of 100, and a score above 68 indicating good usability \citep{brooke1996sus}. 
To evaluate the collected SUS scores, we performed a one-sample t-test to determine if the results were significantly higher than 68. For each SUS sub-score and the customized user experience (UX) questions, we compared the results to a neutral response (score = 3) to confirm the general altitudes of the participants, with the Wilcoxon signed-rank test. When comparing the ratings from participant profiles and UX questions between groups, we used the Mann-Whitney U test. A p-value $<0.05$ was considered to indicate statistical significance for all tests.

For the qualitative data from participants' free-form questionnaires, we reviewed all content and summarized the key insights based on their frequency among participants. These insights provide a better understanding of the semi-quantitative assessments and suggest potential directions for future improvements.

\section{Results} 
\label{sec:results}
Among the participants in the AR group, based on a Likert scale of 1$\sim$5, the level of familiarity with AR (mean$\pm$std) was 2.0$\pm$1.2, indicating some level of familiarity with the technology, while the familiarity with brain anatomy was rated at 3.9$\pm$1.4, which represent low level of familiarity on the topic. For the control group, they are also somewhat unfamiliar with neuroanatomy, with a rating of 4.0$\pm$1.5, which is not statistically different from the AR group (p = 0.87).

\subsection{Educational outcome assessment} 
\label{sec:quiz}

\begin{table}[h!]
\centering
\caption{Comparison of pre-study and post-study quiz scores for AR-based and control groups}
\begin{tabular}{|l|c|c|}
\hline
\multicolumn{3}{|c|}{\textbf{Quiz Scores}} \\ \hline
\textbf{Group} & \textbf{Pre-Study Scores} & \textbf{Post-Study Scores} \\ \hline
\textbf{AR-Based Group} & 40.83\% $\pm$ 7.71\% & 70.28\% $\pm$ 14.11\% \\ \hline
\textbf{Control Group} & 33.89\% $\pm$ 11.24\% & 60.56\% $\pm$ 17.56\% \\ \hline
\end{tabular}

\label{table:quiz_scores}
\end{table}

As shown in Table \ref{table:quiz_scores}, both the AR-based learning group (n = 20) and the control group (n = 20) showed significant improvements in quiz scores before and after the study (p$<$0.0001). The AR group had an average increase of 29.45\%, with pre-study scores of 40.83$\pm$7.71\% and post-study scores of 70.28$\pm$14.11\%. In comparison, the control group had a mean 26.67\% improvement, with pre-study scores of 33.89$\pm$11.24\% and post-study scores of 60.56$\pm$17.56\%. Between the AR-based and control groups, the AR-based group had higher pre-study scores (p = 0.04) while the post-study scores are not significantly different (p = 0.07). Although the AR-based group achieved higher performance gain over the control group on average, the difference is not statistically significant (p = 0.68). 

\subsection{System Usability Scale}
\label{sec:sus}
The overall SUS score calculated based on participant responses for the \textit{NeuroVase} system is 90.0$\pm$5.7, which is significantly higher than the passing score of 68 (p$<$0.001), indicating excellent usability of the \textit{NeuroVase} system. To further inspect the participants' responses to individual SUS sub-questions, we listed their scores in Table \ref{sus_table}. For all sub-questions, we found their responses to be significantly better than the neutral attitude (p$<$0.05).

\begin{table*}[h!]
    \centering
    \caption{System Usability Scale sub-questions and their scores from the AR-based group.}
    \label{sus_table}
    \begin{tabular}{|l|c|}
        \hline
        \textbf{System Usability Scale Sub-questions} & \textbf{Score} \\ \hline
        I think that I would like to use this system frequently. & 4.4 $\pm$ 0.7\\ \hline
        I found this system unnecessarily complex. & 1.3 $\pm$ 0.6\\ \hline
        I thought this system was easy to use. & 4.6 $\pm$ 0.6\\ \hline
        I think that I would need assistance to be able to use this system. & 1.2 $\pm$ 0.5 \\ \hline
        I found the various functions in this system were well integrated. & 4.2 $\pm$ 0.6\\ \hline
        I thought there was too much inconsistency in this system. & 1.5 $\pm$ 0.8\\ \hline
        I would imagine that most people would learn to use this system very quickly. & 4.8 $\pm$ 0.5\\ \hline
        I found this system very cumbersome/awkward to use. & 1.5 $\pm$ 0.8\\ \hline
        I felt very confident using this system. & 4.7 $\pm$ 0.5\\ \hline
        I needed to learn a lot of things before I could get going with this system. & 1.2 $\pm$ 0.4\\ \hline
    \end{tabular}
  
\end{table*}

\subsection{User Experience Assessment}
\label{sec:ux} %new version
\begin{figure}[h!]
\begin{center}
\includegraphics[scale=0.25]{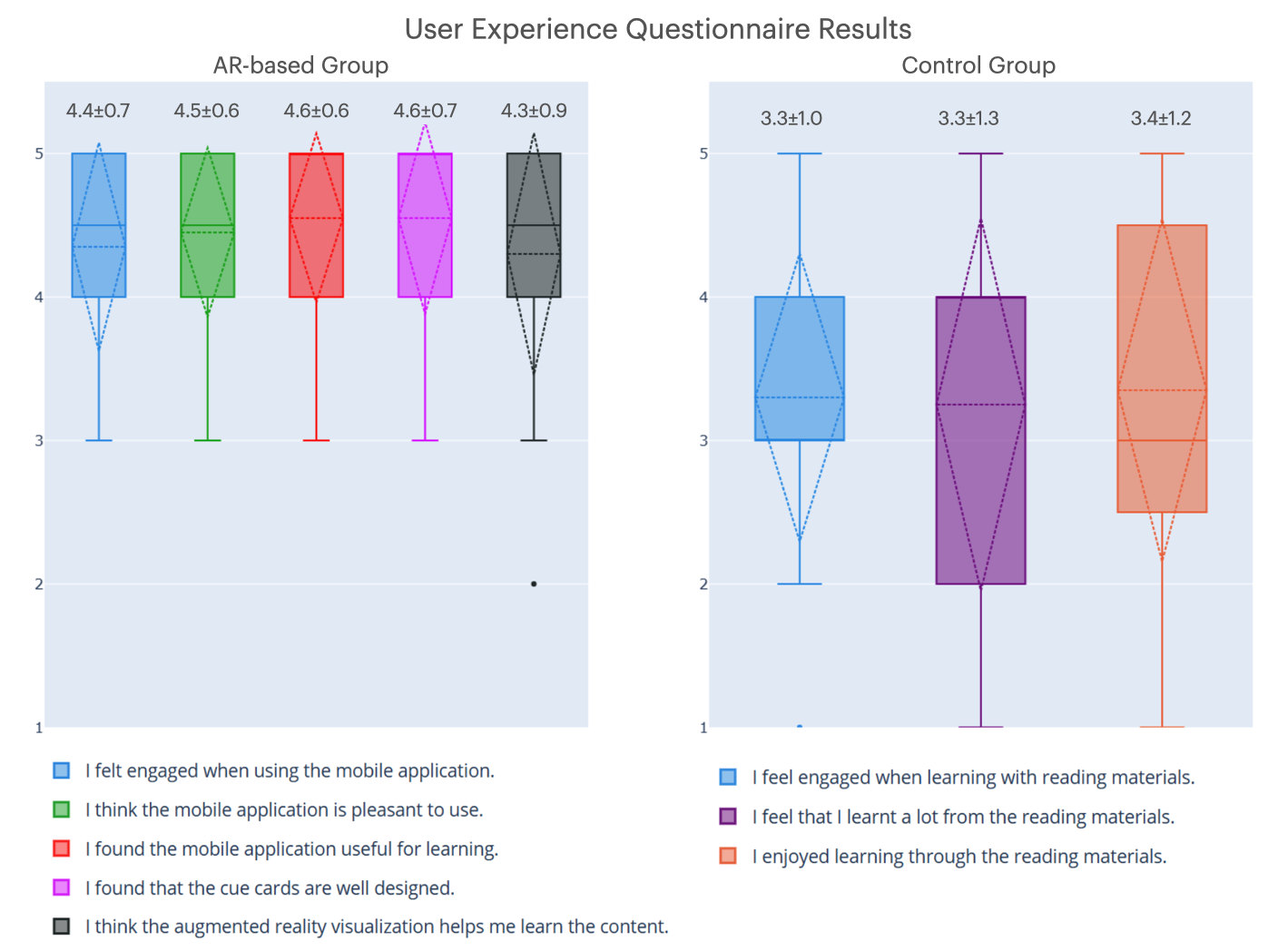}
\end{center}
\caption{Boxplots of Customized UX Questionnaire Scores for the AR and Control Groups.}
\label{fig:uxAR}
\end{figure}

For the AR group, in addition to SUS, five customized UX questions were presented to probe additional dimensions of the user experience concerning the factors related to learning outcomes and the effectiveness of system design elements. Similarly, we presented three UX questions for the control group as well. These questions and the corresponding scores are shown in Figure \ref{fig:uxAR} for the AR-based and the control groups, where a higher score is more desirable. With Wilcoxon signed-rank tests, we confirm that all UX questions for the AR-based group have received more positive responses than the neutral score of 3 (p$<$0.001) while for the control group, the ratings for the three UX questions remain neutral (p$>$0.05). When comparing the enjoyment of two learning methods (``I think the mobile application is pleasant to use" for AR group and ``I enjoyed learning through the reading materials" for control group), AR-based learning is better than traditional paper-based learning (4.5$\pm$0.6 vs. 3.4$\pm$1.2, p = 0.003). In terms of the level of engagement, it is clear that AR-based learning is superior than paper-based learning (4.4$\pm$0.7 vs. 3.3$\pm$1.0, p = 0.001). In terms of perceived usefulness of the \textit{NeuroVase} system, the participants found the mobile AR application a good educational instrument (4.6$\pm$0.6) and the augmented reality visualization facilitate the learning process (4.3$\pm$0.9). As an important component of the full system setup, the participants also rated the overall design of the cue cards highly (4.6$\pm$0.7).

\subsection{Free-form feedback}
For the AR group, we obtained free-form feedback from 19 out of 20 participants to better understand their questionnaire ratings and to gather suggestions for improving the proposed system. For the positive feedback, participants mentioned 3D visualization was helpful, providing better perspectives and view options (11/19), and praised the ease of use (11/19). These echo the high scores for the SUS (90.0$\pm$5.7) and the customized UX question regarding the usefulness of AR visualization in aiding the learning experience (4.3$\pm$0.9). In addition, many participants (7/19) found the user interaction strategy smooth and well designed. On the other hand, some participants noticed that the virtual content triggering mechanism with the cue cards can at times experience glitches (6/19). While a small number of participants (4/19) felt that the educational content was overly detailed, most expressed a desire for additional depths. Specifically, half of the participants suggested more detailed 3D anatomical models and further explanations of some medical jargons to better support lay users. A few also recommended improving content interaction with features, such as the ability to reposition the text box and blood flow animation to enhance understanding of vascular anatomy.

For the control group, we collected feedback from 17 out of 20 participants. Negative feedback mainly focused on content overload, with participants finding the information hard to remember and understand (5/17), and the material overly complex due to detailed descriptions and medical terms (7/17). Many also found the visual representations from textbooks not enough and unclear, suggesting the need for more intuitive medical illustrations (6/17). Some participants reported difficulty staying engaged due to the lack of interactivity (3/17). These comments highlight the need for easy-to-understand learning content, better visual aids, and more engaging interactions.

\section{Discussion}
\label{sec:discussion}
Despite advancements in AR/VR-based neuroanatomy education, several domains remain underexplored. A notable gap lies in the limited focus on vascular anatomy and stroke education, which are essential areas for understanding and treating neurological conditions. Existing applications predominantly address general brain anatomy and connectivity, often with relatively simple learning content. In contrast, \textit{NeuroVase} addresses this gap by providing engaging clinically relevant learning materials tailored for stroke care, developed in collaboration with a clinical specialist.

The \textit{NeuroVase} system distinguishes itself from the mobile AR-based brain stroke education platform proposed by \cite{StrokeEDAR}  in several key ways. First, while their system uses a cube as a physical medium for brain model interaction in AR, \textit{NeuroVase} employs practical cue cards in a dual-model paradigm that offers greater flexibility in managing the learning content. Second, the work by \cite{StrokeEDAR} focuses primarily on patient education, providing only general brain divisions for visualization. In contrast, \textit{NeuroVase} is designed for the education of students and professionals, and thus incorporates more detailed anatomical content, including text-based explanations and imaging data, such as 3D models and MRI overlays. This comprehensive educational approach positions \textit{NeuroVase} as a more versatile and valuable tool for medical education. \textit{Importantly}, our system also features a structured, clinician-informed curriculum comprising three sequential modules, lobar anatomy, vascular anatomy, and arterial territory, designed to build foundational knowledge progressively. This design enhances spatial understanding of the anatomy and facilitates deeper learning, while allowing for future expansion with additional modules built upon the existing framework.

The interface and interaction design of \textit{NeuroVase} were developed to support an intuitive and engaging learning experience in neurovascular anatomy. The use of a tablet-based platform capitalizes on users’ familiarity with and higher accessibility to mobile devices (e.g., tablet), and was well-received as reflected by our questionnaire results.  The customized UX questionnaires further confirmed the perceived usefulness of the proposed AR system, as well as its superior levels of engagement and enjoyment for the proposed AR system over the paper-based learning, which shares identical textual educational content. This echoes observations from previous studies in the benefits of VR/AR-based anatomical learning over textbook and/or lecture-based learning. Notably, the integration of physical cue cards, serving as both AR interaction mediums and standalone study aids, offers flexible learning modalities. These cards also supported structured learning progression, with 91\% of users rating the card design favorably (scores of 4 or 5). The associated interaction design, which allows users to anchor and manipulate a 3D model by repositioning the main cue card and triggering content updates with trigger cards, was appreciated for its hands-on approach and smooth performance (noted by 7/19 participants). Importantly, 12 participants highlighted the value of 3D visualization, and 86\% reported that the interactive 3D model helped them better understand the material. By open-sourcing our developed learning materials on neuroanatomy and stroke, as well as cue card designs, we hope to benefit the development of relevant systems and the community in need. 

Feedback from participants provided valuable insights into areas for improving \textit{NeuroVase}. While the overall usability was rated highly, technical issues such as cue card recognition glitches were reported by 6 out of 19 users. These issues are consistent with prior findings on AR system limitations, particularly with image-based tracking under variable lighting conditions \citep{vuforia}. Future work will explore improving environmental robustness through enhanced card design/production and environment lighting guidelines. Participants also offered specific suggestions for interaction improvements, including enhanced finger-gesture controls (e.g., for rotating models) and the ability to reposition text overlays.  In response to this feedback, we plan to implement a two-finger gesture for more stable model rotation, explore UI elements such as a ``rotation slider” for precision, and redesign cue cards with more distinct visual markers to improve AR tracking reliability. 

In terms of learning outcomes, \textit{NeuroVase} demonstrated strong educational effectiveness. Participants in the AR group showed a higher average improvement of 29.45\%, increasing their quiz scores from 40.83\% to 70.28\%, compared to a 26.67\% increase in the control group, which improved from 33.89\% to 60.56\%. We found that the difference in learning gain between the groups is not statistically significant. This could be partially attributed to the higher pre-quiz scores from the AR-based group (p $<$ 0.05), where the self-rated familiarity with neuroanatomy is slightly better than the control group ($3.9\pm 1.4$ vs. $4.0\pm 1.5$) without a significant difference (p $>$ 0.05). In previous studies on the same topic, the differences in the learning outcomes when comparing AR/VR applications and textbook-based learning have varied. However, the higher engagement and enjoyment from the AR system can offer positive motivation in learning. While both learning methods were effective, these results suggest that the AR system may offer an advantage in enhancing learners’ understanding of neurovascular anatomy. Moreover, users found that the AR visualizations reinforced spatial understanding and memory retention, supporting a more immersive and effective learning experience. Looking forward, we plan to conduct studies with a larger user group, including individuals with varying levels of technological and medical knowledge proficiency, including medical students and emergency care service personnel to evaluate the system’s generalizability across different learner populations.

\section{Conclusion}
\label{sec:conclusion}
In this work, we presented \textit{NeuroVase}, an innovative mobile AR application specifically designed to enhance education of the cerebrovascular system and stroke-related territories. By integrating physical cue cards as tangible interaction tools that also serve as effective standalone review aids, \textit{NeuroVase} bridges the gap between traditional learning and immersive technology. Our user studies demonstrate that the system is not only highly user-friendly and engaging but also effective in improving learners’ understanding. With its rich, interactive 3D visualizations and thoughtfully designed curriculum, \textit{NeuroVase} provides a powerful and transformative supplement to conventional educational methods.

\section*{Conflict of Interest Statement}
%All financial, commercial or other relationships that might be perceived by the academic community as representing a potential conflict of interest must be disclosed. If no such relationship exists, authors will be asked to confirm the following statement: 

The authors declare that the research was conducted in the absence of any commercial or financial relationships that could be construed as a potential conflict of interest.

\section*{Author Contributions}
B.J. was responsible for the study concept, image processing, educational material creation, software implementation, experiment execution, data analysis and interpretation, and writing of the manuscript. M.L.K was responsible for educational material creation and editing of the manuscript. M. K. was responsible for overall study oversight and editing of the manuscript. Y.X. was responsible for the study concept, image processing, data analysis, writing and editing of the manuscript, and provided the overall study oversight. All authors contributed to the article and approved the submitted version.

%The Author Contributions section is mandatory for all articles, including articles by sole authors. If an appropriate statement is not provided on submission, a standard one will be inserted during the production process. The Author Contributions statement must describe the contributions of individual authors referred to by their initials and, in doing so, all authors agree to be accountable for the content of the work. Please see  \href{https://www.frontiersin.org/about/policies-and-publication-ethics#AuthorshipAuthorResponsibilities}{here} for full authorship criteria.

\section*{Funding}
The study was supported by the Fonds de recherche du Qu\'ebec Nature and Technologies (FRQNT), Team Research Project Grant: 2022-PR296459; Fonds de recherche du Qu\'ebec-Sant\'e Junior 1 Research Scholar program, and Parkinson Qu\'ebec.

%\section*{Acknowledgments}
%This is a short text to acknowledge the contributions of specific colleagues, institutions, or agencies that aided the efforts of the authors.

%\section*{Supplemental Data}
% \href{http://home.frontiersin.org/about/author-guidelines#SupplementaryMaterial}{Supplementary Material} should be uploaded separately on submission, if there are Supplementary Figures, please include the caption in the same file as the figure. LaTeX Supplementary Material templates can be found in the Frontiers LaTeX folder.

\section*{Data Availability Statement}
The datasets generated for this study can be found in the (OSF) repository: \href{https://osf.io/7bq2s/overview?view_only=19cfc4c64ea747d089b841f0fe1ddd32}{\textit{https://osf.io/7bq2s}}.

%https://osf.io/7bq2s/overview?view_only=19cfc4c64ea747d089b841f0fe1ddd32.
% Please see the availability of data guidelines for more information, at https://www.frontiersin.org/about/author-guidelines#AvailabilityofData

\bibliographystyle{Frontiers-Harvard} %  Many Frontiers journals use the Harvard referencing system (Author-date), to find the style and resources for the journal you are submitting to: https://zendesk.frontiersin.org/hc/en-us/articles/360017860337-Frontiers-Reference-Styles-by-Journal. For Humanities and Social Sciences articles please include page numbers in the in-text citations 
\bibliography{test}

%%% Make sure to upload the bib file along with the tex file and PDF
%%% Please see the test.bib file for some examples of references

\clearpage

%\section*{Figure captions}

%%% Please be aware that for original research articles we only permit a combined number of 15 figures and tables, one figure with multiple subfigures will count as only one figure.
%%% Use this if adding the figures directly in the mansucript, if so, please remember to also upload the files when submitting your article
%%% There is no need for adding the file termination, as long as you indicate where the file is saved. In the examples below the files (logo1.eps and logos.eps) are in the Frontiers LaTeX folder
%%% If using *.tif files convert them to .jpg or .png
%%%  NB logo1.eps is required in the path in order to correctly compile front page header %%%

%\clearpage

%\section*{Table captions}

%%% If you don't add the figures in the LaTeX files, please upload them when submitting the article.
%%% Frontiers will add the figures at the end of the provisional pdf automatically
%%% The use of LaTeX coding to draw Diagrams/Figures/Structures should be avoided. They should be external callouts including graphics.

\end{document}